# Experimental Observation of Invariance of Spectral Degree of Coherence with Change in Bandwidth of Light

Bhaskar Kanseri* and Hem Chandra Kandpal

Optical Radiation Standards, National Physical Laboratory, New Delhi-110012 India

*Corresponding author: kanserib@mail.nplindia.ernet.in

An experimental study is conducted to show the effect of the change in bandwidth of light on the spectral degree of coherence at a pair of points in the cross-section of a beam. For this purpose a polychromatic source and a monochromator with variable entrance and exit slits were used to produce a variable bandwidth source. The classic Young's interferometer was used to produce an interference pattern. The spectral measurements of the visibility of the interference fringes show that the spectral degree of coherence remains unaffected by the change in the frequency pass-band of the light.

In the past several decades, partially coherent light was investigated theoretically and experimentally with great interest [1]. The central quantities in these studies were the complex degree of coherences both in space-time [2] and in space-frequency domain [3]. The effects of spatial correlations (source correlations) on the spectral properties of light were studied in details and many applications were suggested [4]. In recent investigations, the role of polarization became apparent in connection with the coherence phenomenon and a new theory called the unified theory of coherence and polarization [5] was proposed to resolve many problems of optics [6]. However, the scalar coherence theories have great potential for determining various source and field dependent properties [1].

In space-frequency domain, the quantification of coherence is done by the degree of spectral coherence. One might expect that the degree of spectral coherence increases when identical filters centred at the same frequency, but having narrow band pass are put before both the slits of the interferometer. The theoretical study of this effect made by Wolf [7] in 1983 concludes otherwise. One of the important results of the paper was that *the degree of spectral coherence remains unchanged by filtering*. Still no experimental study of this aspect is found in literature, though in the recent past, a theoretical study showing the relation between space-time and space-frequency complex degree of coherences was reported [8].

In this letter it is shown qualitatively and quantitatively that the spectral degree of coherence does not change with the change in bandwidth of the source. Instead of using a tunable interference filter as a secondary source, we have used another scheme. A variable bandwidth source is constructed by using a variable slit monochromator placed in line with a polychromatic source. The bandwidth of the secondary source is altered by changing simultaneously the entrance and the exit slit of the monochromator by the same amount. Using a double-slit after the synthesized secondary source, the interference fringes were obtained on an observation plane. The spectral visibility of the fringes with respect to the change in bandwidth of light was measured and was found nearly constant. This invariance of fringe contrast (spectral) is consistent with the theoretical approach given in ref. [7]. This experimental work, we believe is the first experimental observation of the invariance of degree of spectral coherence with change in bandwidth of the light.

In the Young's interference experiment, the absolute value of the spectral degree of coherence $|\mu(\mathbf{r}_1,\mathbf{r}_2,\omega)|$, also known as spectral visibility $v(\mathbf{r})$ is a direct measure of the contrast of the interference fringes formed using quasimonochromatic light. In case, the spectral densities at the observation point due to separate slits are approximately the same, one can readily find [1]

$$v(\mathbf{r}) = |\mu(\mathbf{r}_1,\mathbf{r}_2,\omega)| = \frac{S_{max} - S_{min}}{S_{max} + S_{min}} \tag{1}$$

where $S_{max}$ and $S_{max}$ are respectively the maximum and the minimum values of spectral density around the observation point. The cross-spectral density of light is given by [1]

$$W(\mathbf{r}_1,\mathbf{r}_2,\omega) = \langle U^*(\mathbf{r}_1,\omega)U(\mathbf{r}_2,\omega)\rangle, \tag{2}$$

where $U(\mathbf{r}_i,\omega)\exp(-i\omega t)$ denotes an ensemble of the monochromatic oscillations and the angular brackets show the ensemble average.

It was shown by Wolf [7] that in case two identical filters having amplitude transmission function $T(\omega)$ are put after the double-slit, covering both the slits, then the light emerging out of the filters is given by the ensemble $T(\omega)U(\mathbf{r}_i,\omega)\exp(-i\omega t)$. So using Eq. (2), the cross-spectral density of the filtered light, taking the factor $T^*(\omega)T(\omega)$ outside the averaging, will be given by [7]

$$W^+(\mathbf{r}_1,\mathbf{r}_2,\omega) = |T(\omega)|^2 W(\mathbf{r}_1,\mathbf{r}_2,\omega). \tag{3}$$

If we assume that the cross-spectral density $W(\mathbf{r}_1,\mathbf{r}_2,\omega)$ is a continuous function of frequency $\omega$, and $\Delta\omega$ is so small that $W(\mathbf{r}_1,\mathbf{r}_2,\omega)$ does not appreciably change across the effective pass-band $\omega_0 - \frac{1}{2}\Delta\omega \ll \omega \ll \omega_0 + \frac{1}{2}\Delta\omega$ of the filters, then we can replace $W(\mathbf{r}_1,\mathbf{r}_2,\omega)$ by $W^+(\mathbf{r}_1,\mathbf{r}_2,\omega)$ in Eq. (2). The complex degree of spectral coherence for the unfiltered light of central frequency $\omega_0$ is given by [1, 7]

$$\mu(\mathbf{r}_1,\mathbf{r}_2,\omega_0) = \frac{W(\mathbf{r}_1,\mathbf{r}_2,\omega_0)}{[W(\mathbf{r}_1,\mathbf{r}_1,\omega_0)]^{1/2}[W(\mathbf{r}_2,\mathbf{r}_2,\omega_0)]^{1/2}}. \tag{4}$$

In a similar way, the complex degree of spectral coherence at frequency $\omega_0$ of the filtered light at the two points is written as [7]

$$\mu^+(\mathbf{r}_1,\mathbf{r}_2,\omega_0) = \frac{W^+(\mathbf{r}_1,\mathbf{r}_2,\omega_0)}{[W^+(\mathbf{r}_1,\mathbf{r}_1,\omega_0)]^{1/2}[W^+(\mathbf{r}_2,\mathbf{r}_2,\omega_0)]^{1/2}}. \tag{5}$$

Using Eq. (3) with (4) and (5), we get [7]

$$\mu^+(\mathbf{r}_1, \mathbf{r}_2, \omega_0) = \mu(\mathbf{r}_1, \mathbf{r}_2, \omega_0), \tag{6}$$

i.e. the complex degree of spectral coherence remains unchanged by filtering. From Eq. (6), we can readily see that the absolute values of these quantities, which can directly be calculated using Eq. (1) by measuring experimentally the spectral densities, will also be equal.

The experimental arrangement is shown in Fig. 1. A tungsten-halogen lamp (Mazda, colour temperature = 3200 K), operated at 700 W using a regulated dc power supply (Heinzinger, stability 1 part of $10^4$) was utilized as a polychromatic continuous spectrum source. The outer glass jacket of the lamp was diffused to obtain uniform illumination. The light coming out from the source was passed through a microprocessor controlled monochromator (CVI, Digikrom) having variable entrance and exit slits. The beam emerging from the exit slit of the monochromator was made incident on a single-slit (slit width 500 µm) placed at a distance 40 cm from the exit slit. The beam emerging from the single-slit illuminated a double-slit plane having rectangular slits with slit width 150 µm and slit separation 350 µm and placed 80 cm away from the single-slit. Interference fringes were obtained at the observation plane at a distance of 40 cm from the double-slit. The spectral measurements were carried out using a fibre-coupled spectrometer (Photon Control, SPM002) interfaced with a personal computer.

The central wavelength of the output beam was selected at 555 nm. This wavelength was chosen due to its maximum sensitivity for human eye providing convenience in aligning the optics. Both the slits of the monochromator were opened by the same width so that the output beam spectrum could remain centred at the desired wavelength [9, 10]. By changing the width of both the slits by known amount, the bandwidth of the spectra was changed. Fig. 2 shows the graphical representation of the light spectra obtained for different slit widths. We observe that by increasing the slit opening, the bandwidth of the beam could be increased and vice-versa. However, for smaller slit widths, the transmitted light intensity also reduces which demand sensitive detectors for measurement. The use of monochromator provided operational flexibility.

Partial coherence develops when radiation propagates, even in free space. In far zone, the diameter of the region outside which there is complete incoherence is given by van-Cittert Zernike theorem [1] and is proportional to $\frac{\bar{\lambda} R}{\rho}$ [11] for a rectangular slit, where $\bar{\lambda}$ is the central wavelength, $\rho$ is the slit width and R is the distance of the observation plane from the slit. Thus the exit slit width of the monochromator plays a role of deterministic slit width for the developed coherence region at the double-slit plane. This effect is evident from the photograph of the double-slit diffraction patterns for different slit widths of the monochromator, as shown in Fig. 3. We see that for higher slit opening, the coherence developed at double-slit plane is low, resulting to suppression of the interference effect. When the slit spacing is decreased, according to van-Cittert Zernike theorem, the coherence developed at the double-slit plane increases and the interference effects start dominating. For very small slit separation, only the interference effects remain (Fig. 3 e). This feature confirms the role of van-Cittert Zernike theorem which establishes an inverse relation between the slit width and the developed degree of coherence at any observation plane. To observe the behaviour of the degree of spectral coherence with the bandwidth, the coherence at double-slit should not be influenced by the other factors, namely slit width etc. However, this is not the case in this study.

In this experiment, to achieve this feat, we modified the setup by introducing a single-slit having constant slit width in between the monochromator and the double-slit (Fig. 1). This additional slit isolates the effect of slit width variation on the spatial coherence region developed at the double-slit plane. Thus the spatial coherence region developed at the double-slit plane does not effectively change throughout the experiment and remains invariant with the geometrical conditions of the monochromator slits. However, introduction of the single-slit does not appreciably affect the bandwidth of the emerging light. A quantitative representation of the change in bandwidth of light with the slit width of the monochromator for the modified set up, measured at the double-slit plane using the fibre-coupled spectrometer is shown in Fig. 4. As evident from the figure, the resultant bandwidth reduces slightly; however, the dependence of bandwidth on the slit width remains linear. Thus the use of monochromator with the single-slit plays the role of a tunable filter mentioned in Ref. [7].

The photographs of interference fringes obtained by decreasing the monochromator slit widths, i.e. the bandwidth of light, are shown in Fig. 5. A careful look on the interference fringes makes it apparent that there is a very small change in the visibility of the fringes with the change in bandwidth, though the intensity of the fringes reduces rapidly with decrease in bandwidth. The observed little increment in fringe visibility is due to the fact that the complex degree of coherence (time domain) undergoes slight enhancement with the decrease in bandwidth of light [7]. It is worth to recall that the effective bandwidth of the light emerging out of the monochromator, i.e. $3\times10^{12}$ Hz was so narrow ($3\times10^{12}$ Hz $\ll 5\times10^{14}$ Hz, peak frequency) that both the modulus and the phase of $W(\mathbf{r}_1,\mathbf{r}_2,\omega)$ and the spectral densities $S_1(\mathbf{r})$ and $S_2(\mathbf{r})$ for the pair of points could be assumed to be substantially constant.

To validate our findings, the spectral measurements were made using the spectrometer by tracing the fiber tip horizontally across the interference fringes. The spectral densities due to individual slits were measured at the axial point and were found the same, i.e. $S_1(\mathbf{r}) \approx S_2(\mathbf{r})$. The spectral visibility of the fringes was determined using Eq. (1). A plot of absolute value of degree of spectral coherence with the change in bandwidth of light is shown in Fig. 6. The slit width scale was changed into the bandwidth scale using the graphical equivalence shown in Fig. 4. We observe that the degree of spectral coherence remains nearly constant throughout the bandwidth range. The higher variance in the repeat measurements for low bandwidth regime may be due to the limited sensitivity of the spectrometer at low intensity levels.

In nutshell, the behaviour of the spectral degree of coherence is studied experimentally with the change in the bandwidth of the light. It is found that like the theoretical outcome, this quantity remains invariant with these bandwidth variations in the source spectra. This experimental finding not only provides verification to the theoretical study of invariance of degree of spectral coherence by filtering of light but also paves way for finding relation experimentally between the time dependent and the frequency dependent coherence functions.

We express our thanks to the Director, National Physical Laboratory, New Delhi, India for permission to publish this paper. Author BK wishes to thank the Council of Scientific and Industrial Research (CSIR), India for financial support as Senior Research Fellowship.

**Caption for figures**

**Fig. 1:** Schematic of the experimental setup. Notations, S a continuous spectrum polychromatic source, D diffuser, M monochromator with microprocessor (MP) control, SS single-slit, DS double-slit, R observation plane, SM spectrometer with fibre (F) coupling and DP personal computer. In monochromator (M), 1 and 2 represent the entrance slit and exit slit, respectively.

**Fig. 2:** Change in the spectral density and bandwidth of light measured at plane R for varying slit width of the monochromator. Curves are plotted for slit width (i) 100 µm, (ii) 500 µm, (iii) 1000 µm, (iv) 1500 µm and (v) 2000 µm. Dots show the experimental data points connected by the solid line.

**Fig. 3:** Photographs showing the two-slit diffraction patterns for different slit widths (a) 2000 µm (b) 1500 µm (c) 1000 µm (d) 500 µm and (e) 100 µm, when the single-slit is removed from the experimental arrangement shown in Fig. 1. The contrast in the fringes increases with decrease in slit width.

**Fig. 4:** Graphical representation of the full width at half maxima (FWHM) of the spectra (after introducing the single-slit) plotted with respect to the slit width of the monochromator. Dots represent the experimental data points while the trend line shows the linear fit. The error bars illustrate the variance for the repeat measurements.

**Fig. 5:** Photographs showing the double-slit interference fringe for different slit widths. The slit widths are (a) 2000 µm (b) 1500 µm (c) 1000 µm (d) 500 µm (e) 100 µm. The contrast in the interference fringes appears to be nearly same for all these conditions.

**Fig. 6:** Graph showing the change in absolute value of the degree of spectral coherence (spectral visibility) for a pair of points in the double-slit plane with the change in bandwidth of light. The dashed line is the trend line showing linear fit.

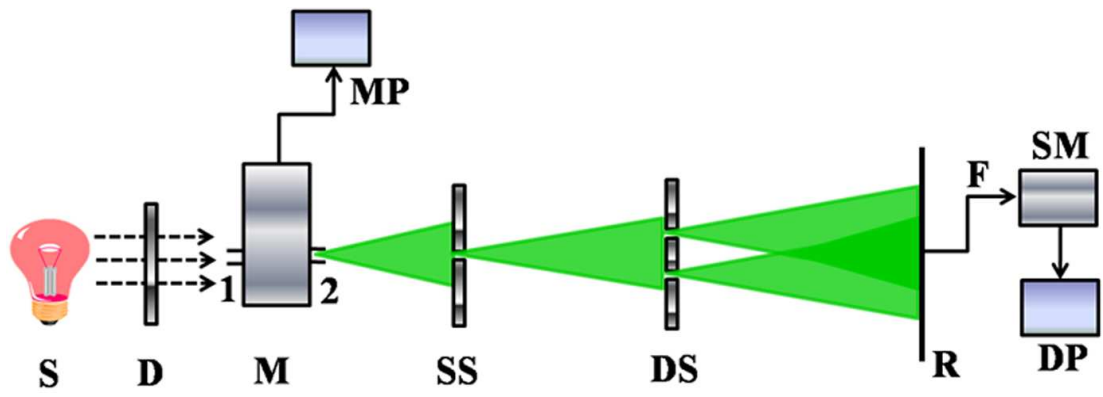

**Fig. 1**

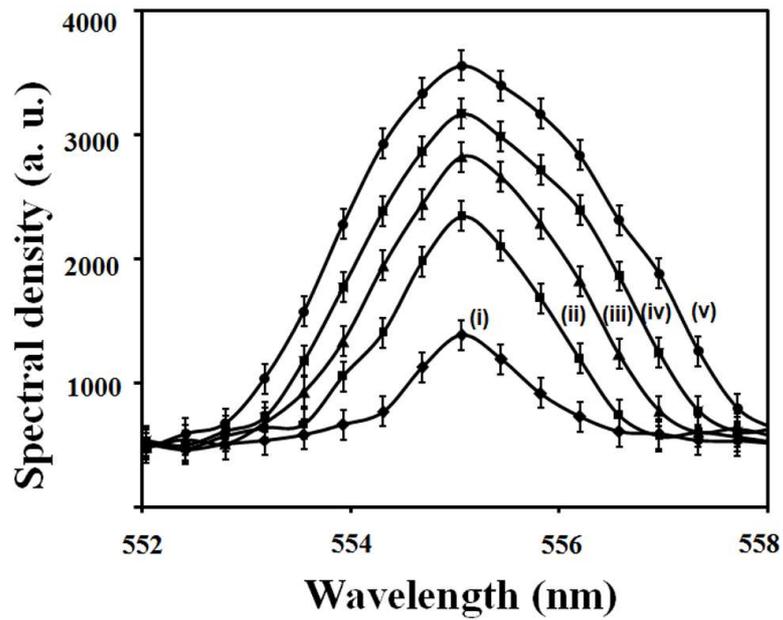

**Fig. 2**

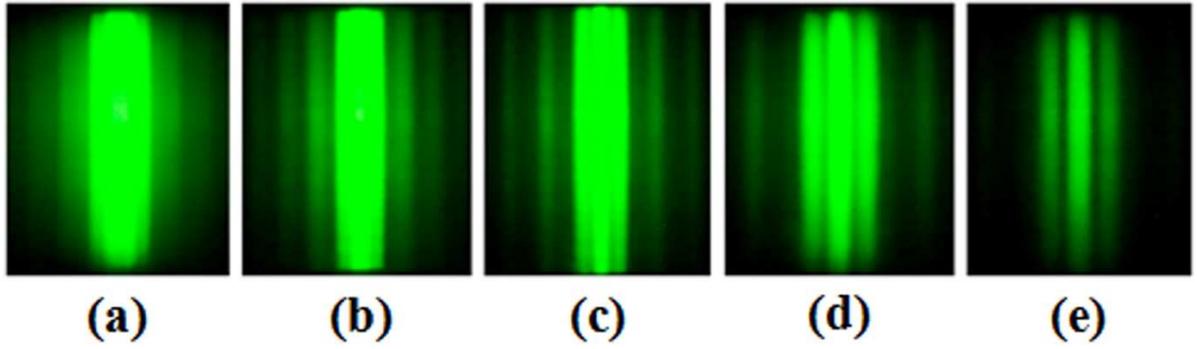

**Fig. 3**

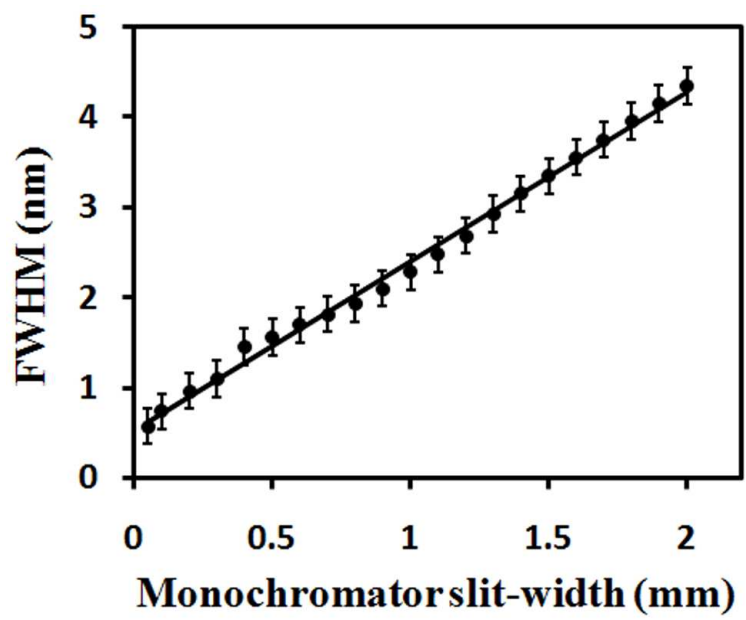

**Fig. 4**

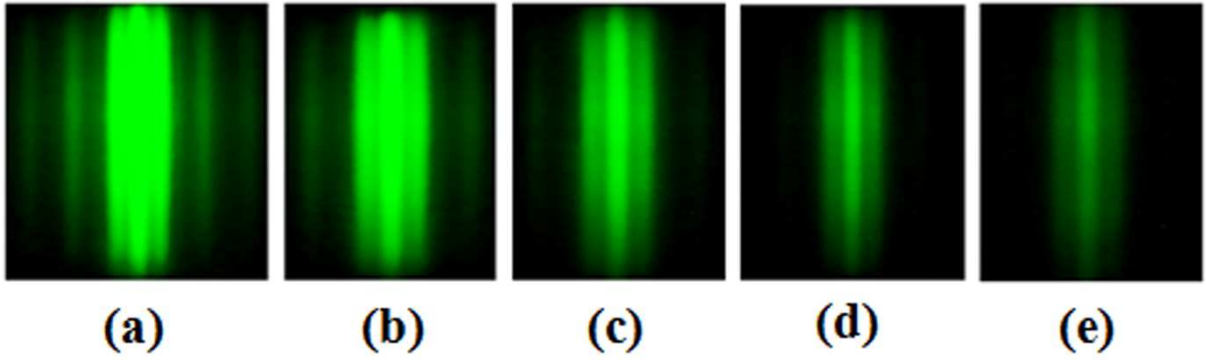

**Fig. 5**

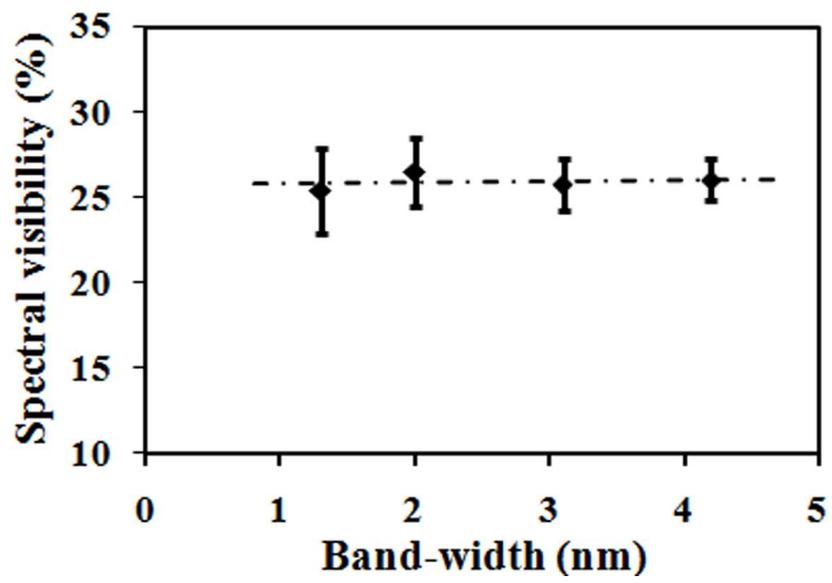

**Fig. 6**